# Contact melting and the structure of binary eutectic near the eutectic point

Bystrenko O.V.[1,2] and Kartuzov V.V.[1]

[1] Frantsevich Institute for material science problems, Kiev, Ukraine,
[2] Bogolyubov Institute for theoretical physics, Kiev, Ukraine

**Abstract**. Computer simulations of contact melting and associated interfacial phenomena in binary eutectic systems were performed on the basis of the standard phase-field model with miscibility gap in solid state. It is shown that the model predicts the existence of equilibrium three-phase (solid-liquid-solid) states above the eutectic temperature, which suggest the explanation of the phenomenon of phase separation in liquid eutectic observed in experiments. The results of simulations provide the interpretation for the phenomena of contact melting and formation of diffusion zone observed in the experiments with binary metal-silicon systems.

1. **Introduction.**

Phenomenon of contact (eutectic) melting (CM) is rather common for multicomponent systems and has important industrial implications [1], which motivate its experimental and theoretical studies. The aim of this work is a theoretical study of the properties of CM in binary eutectic systems and the interpretation of the results of recent experiments particularly focused on the investigation of the interfacial phenomena associated with CM [2, 3]. In these experiments, the binary systems consisting of metal Au, Al, Ag, and Cu particles (of $\approx 5 \cdot 10^{-6} m$ size) placed on amorphous or crystalline silicon films were examined. The most general conclusions obtained in these works can be formulated as follows: 1) for the temperatures below the eutectic point, under isothermal conditions, the solid-solid interface forms a diffusion zone (DZ) with typical width of $\approx 10^{-5} m$, which may reduce in size for lower temperatures; 2) for the temperatures within the range $10 - 20°C$ above the eutectic point, under isothermal conditions, the formation of the liquid eutectic layer at the interface (i.e., the phenomenon of CM) is observed, which is, however, always preceded by the formation of the DZ. Notice that the above experimental results appear to be nontrivial. The rates of melting are typically much higher than the mass transport processes in solids; therefore, for the temperatures above the eutectic point, the metal particles should completely melt prior to the formation of DZ. The shrinking of the DZ with lowering the temperature observed in some experiments looks unexpected as well, since the entropy-driven diffusive processes would normally result in increase of the DZ. The processes occuring in the DZ in the above mentioned experiments are very complicated due to the formation of a large number of compounds with different structure and thermodynamical properties. However, one should expect that the most general features of these phenomena would be reproduced in the theoretical approaches based on the general thermodynamical assumptions. For this reason, we took as a tool for our study the standard version of phase-field theory (PFT) [4-6], which contains the basic thermodynamics of binary systems with eutectic properties being the result of immiscibility of components in solid state. In addition, the investigation of interfacial phenomena associated with CM on the basis of the standard version of PFT is of interest for theory, since, at present, no such studies have been reported.

2. **Formulation of the model.**

Within the PFT, the dissipative kinetics of a system is described in terms of phase-field variables. The relevant kinetic equations for an isothermal binary system can be obtained from the free energy functional,

$$F = \int \left[ f(\varphi,c,T) + \frac{\varepsilon_C^2}{2}(\nabla c)^2 + \frac{\varepsilon_\Phi^2}{2}(\nabla \varphi)^2 \right] dV, \qquad (1)$$

where $f(\varphi, c, T)$ is the free energy density; $T$, $c$, and $\varphi$ denote the temperature, concentration and the phase field variable, respectively. The phase-field variable specifies the state, $\varphi = 0$ for solid, and $\varphi = 1$ for liquid phase. By applying the basic principles of thermodynamics, one obtains the PFT-equations

$$\frac{\partial \varphi}{\partial t} = -M_\Phi \left[ \frac{\partial f}{\partial \varphi} - \varepsilon_\Phi^2 \nabla^2 \varphi \right] \qquad (2)$$

$$\frac{\partial c}{\partial t} = \nabla \cdot \left[ M_C c(1-c) \nabla \left( \frac{\partial f}{\partial c} - \varepsilon_C^2 \nabla^2 c \right) \right]. \qquad (3)$$

Here the quantities $M_\Phi$ and $M_C$ are the kinetic coefficients specifying the relaxation rate of the system to the equilibrium; $\varepsilon_\Phi$ and $\varepsilon_C$ are the parameters responsible for the surface energy of solid-liquid and solid-solid interfaces.

The specific features of a system are determined by the particular form of free energy density $f(\varphi,c,T)$. In this work we use the standard expression for a non-ideal binary system [5, 6]

$$\begin{aligned} f(\varphi,c,T) = (1-c)f_A(\varphi,T) + cf_B(\varphi,T) + \frac{RT}{v_m}[c\ln c + (1-c)\ln(1-c)] \\ + c(1-c)[\Omega_S[1 - p(\varphi)] + \Omega_L p(\varphi)] \end{aligned}, \qquad (4)$$

where $R$ is the gas constant, $v_m$ is the molar volume, $\Omega_S$ and $\Omega_L$ are the mixing energies for the solid and liquid state, respectively. In what follows, we assume the mixing energy in the liquid state $\Omega_L = 0$, thus, the eutectic properties of the system (miscibility gap) are associated with the non-zero mixing energy $\Omega_S = \Omega$ in the solid state only. The component free energies $f_A(\varphi,T)$ and $f_B(\varphi,T)$ have the form

$$f_{A,B}(\varphi,T) = W_{A,B} g(\varphi) + L_{A,B} \frac{T_M^{A,B} - T}{T_M^{A,B}} p(\varphi), \qquad (5)$$

where $T_M^{A,B}$ are the melting temperatures, $W_{A,B}$ are the energy barriers associated with the surface energy of liquid-solid interfaces, $L_{A,B}$ are the component latent heats for the components A and B; the functions $g(\varphi) = \varphi^2(1-\varphi)^2$ and $p(\varphi) = \varphi^3(10 - 15\varphi + 6\varphi^2)$ are the model barrier and interpolating functions constructed in such a way as to provide the description of the liquid-solid interfaces of a finite width. The kinetic coefficients $M_\Phi$ and $M_C$ are determined for a binary system as

$$M_\Phi = (1-c)M_A + cM_B \qquad (6)$$

$$M_C = \frac{D_S + p(\varphi)(D_L - D_S)}{RT/v_m} \qquad (7)$$

with $D_S$ and $D_L$ being the solid and liquid diffusivities, respectively.

It should be noted that the relevance of the above model for the description of eutectic systems (in particular, for the systems with different component structure) has been questioned in view of the fact that it yields zero surface energy in the sharp-interface limit and because it ignores the difference in crystalline structure of the components [6] . In this connection, the following comment should be made. It is obvious that the formation of DZ in binary eutectic systems observed in experiments is the process of the relaxation of the system from the initial state of complete segregation of components to the thermal equilibrium. Since in binary eutectic this latter is expected to be the state of complete decomposition as well, it is natural to assume that the DZ can be regarded as the diffuse solid-solid interface associated with the equilibrium in the system. This is the basic assumption underlying all the following consideration. This means, in particular, that quite similar DZ can be obtained alternatively as the result of complete equilibration in a binary eutectic system with arbitrary initial composition profiles. An additional argument in favor of such assumption is the above mentioned shrinking of DZ with lowering the temperature observed in experiments with Al-Si systems [3]. As the result, the thickness of the interface (as well as the associated surface energy) becomes a real physical quantity (but not the model parameter as in Ref. [6]), which should be measured in experiment. Furthermore, we neglect the effects of a large number of possible compounds in DZ with different structural and thermodynamical properties, and use the concentration field as the only variable describing the composition with indefinite (amorphous or small-grained) structure. As we shall see below, such a simplified approach reproduces a number of phenomena to be examined.

3. **The results of numerical simulations.**

To be specific, we give here the results of simulations for Cu-Si system. In PFT simulations, the following parameters were used ($A \equiv Cu; B \equiv Si$): the domain of solution $X_m = 2.0 \cdot 10^{-5}$ m; time interval $t_{max} = 4000\ sec$; melting temperatures $T_M^A =$ 1357 °K and $T_M^B = 1685$ °K; latent heats $L_A = 1.837 \cdot 10^9\ J/m^3$ and $L_B = 4.488 \cdot 10^9\ J/m^3$; molar volume $v_m = 8.57 \cdot 10^{-6}\ m^3/mole$; surface energies for liquid/solid interfaces $\sigma_A = \sigma_B = 0.3\ J/m^2$; diffusivities $D_S = 10^{-13}\ m^2/sec$ and $D_L = 10^{-9}\ m^2/sec$; kinetic parameters for interfaces $\mu_A = \mu_B = 5 \cdot 10^{-5}\ m/(°K sec)$; mixing energy for solid phase $\Omega = 3.635 \cdot 10^9\ J/m^3$.

Most of the above parameters are the experimental data available for Cu-Si system or the typical numbers for the relevant quantities (like solid/liquid diffusivities) used in similar simulations [7-9]. The experimental value for the important quantity $\Omega$ is not available in the literature; we fitted it to make the eutectic temperature match its experimental value $T_{eut} = 1075°K$. In order to make computer simulations less time consuming, the kinetic coefficients $\mu_{A,B}$ are set much less than the typical values used in similar PFT simulations [9]. However, it is to emphasize that the simulations performed were aimed at the investigation of the steady (equilibrium or metastable) states, and, as is seen from Eq. (2), the kinetic coefficients affect the rate of the dissipative relaxation of the system rather than the final steady states. The important for the model quantity, the width of the DZ, is determined by the concentration gradient coefficient $\varepsilon_C$. The latter was fitted as well to provide the typical value $d \approx 4 \cdot 10^{-6}\ m$ observed experiments. The associated interfacial energy of DZ evaluated in simulations was on the order of $\Sigma \approx 500 \, J/m^2$.

The above parameters $\sigma_{A,B}$ and $\mu_{A,B}$ are needed to calculate the phase gradient coefficient $\varepsilon_\Phi$, the barrier energy $W_{A,B}$, and the kinetic parameters $M_{A,B}$ entering the PFT equations; for this purpose, the standard relations given in Ref. [5] were employed. The model parameter specifying the liquid-solid interface width was set $\delta/X_{max} = 0.01$. Numerical solution of Eqs. (2-3) was based on finite-element scheme, by using the open-source FEniCS libraries [10]. The absolute error for composition and phase profiles obtained in computations was kept less than $10^{-5}$.

A part of the experimental phase diagram for Cu-Si near the eutectic point [7] is displayed in Fig. 1. To reproduce the results of experiments with the formation of the DZ [2], we performed PFT simulations in 1D (planar) geometry for the points D1 ($T_{D1} = 923\,°K\,(650\,°C)$) and D2 ($T_{D2} = 1083\,°K\,(810\,°C)$) for the eutectic concentration c=0.3. For this purpose Eqs. (2-4) were solved with initial conditions describing a contact of two solid components (Fig. 2, line 1).

Notice, that the PFT equations allow overheated metastable solid state $\varphi = 0$ to exist infinitely long. For this reason, to trigger the kinetic processes, the initial phase distribution was modified to include 0.05-0.15 - amplitude random noise. In principle, the effects of random fluctuations can be taken into account by adding noise source to PFT equations or using the renormalization approach [11-13]. However, we did not consider the fluctuations in simulations, since these latter affect the kinetics rather than the final steady states.

The results of simulations for the temperatures specified in experiments with Cu-Si system [2] are displayed in Figs. 3, 4. For the temperature $T_{D1} = 923\,°K\,(650\,°C)$ below the eutectic point (Fig. 3), the simulations show that, as the result of the relaxation, there forms a final steady state corresponding to 'solid A + diffuse interface + solid B' composition profile. It is worth mentioning that using any of the initial composition distributions given in Fig. 2 yields identical final steady-state composition profiles, which is the result of the equilibrium nature of this state. The area in grey (diffuse interface) in Fig. 2 is interpreted as a DZ. Another important observation obtained in PFT simulations is the reduce in the width of DZ with lowering the temperature.

The simulations performed for the temperature $T_{D2} = 1083\,°K\,(810\,°C)$ above the eutectic point (Fig. 4) indicate the formation of a liquid area within the steady DZ, quite consistent with experimental observations [2]. Notice that the steady three-phase (solid A + liquid + solid B) profiles displayed in Fig. 4 can be viewed as the state of complete decomposition (like those shown in Fig. 3), but with a liquid interface. It is to mention that similar simulations performed for the parameters specific for Ag-Si system yielded quite analogous results.

The fundamental question, which arises in this context, is that of the nature of the steady states observed, i.e., if they are metastable or equilibrium ones. To address this question, we performed series of runs with various initial conditions in the vicinity of the expected eutectic point with monitoring the free energy of the system defined by Eq. (1). Depending on the initial phase and concentration distributions, for the same temperature and average concentration (i.e., for the same point on the phase diagram), the system may be trapped in a metastable state, or proceed to the thermal equilibrium. An example of the behavior of free energy during relaxation for the temperature $T_{D2} = 1083\,°K\,(810\,°C)$ and eutectic concentration $c = 0.3$ (point D2) for the three-phase (solid A + liquid + solid B) and two-phase (solid A + solid B) states is given in Fig. 5 (with free energy density measured in units of $f_0 = RT_A^M/v_m = 1.317 \cdot 10^9\ J/m^3$). The free energy density for liquid state evaluated for the same point on the phase diagram is much higher, $f_L/f_0 = 0.0705$. Therefore, one must conclude, that 1) the liquid state at the point D2 is metastable; 2) thermal equilibrium in the system is associated with the steady three-phase 'solid A + liquid + solid B' state (Fig. 4), because it has the lowest free energy. It is to emphasize, that this theoretical (based on PFT) conclusion, while being consistent with above cited experiments, contradicts the experimental phase diagram (Fig. 1), since the point D2 lies here within the area of liquid phase. To examine this issue in more detail, we performed further simulations aimed at the evaluation of free energy of possible steady states in the vicinity of eutectic point and deriving the associated phase diagram. The results are given in Figs. 6-8.

The most unexpected finding is that, according to the PFT model used, the solidus line $T = T_{eut}$ turns out to have a 'fine structure', i.e., it splits and expands to the relatively narrow (about $\approx 50\,°C$ width) area above the eutectic temperature, where the above mentioned states describing the complete decomposition with the liquid interface ('solid A + liquid + solid B') are thermodynamically stable. The point D2 lies therewith within this area, which explains the experimental observations of the DZ with the layer of liquid eutectic [2].

In this context a few words should be said about the standard technique of the common tangent construction for obtaining phase diagrams. The latter ignores the gradient terms in the free energy functional (1) and, therefore, implies the equilibrium in the system to be associated with the uniform concentration and the phase field. As the result, the contribution to the free energy of the interfaces is neglected and the description of the complicated interfaces like given in Fig. 4 with high interfacial energy is impossible. Notice that the profiles given in Fig. 4 are the exact solutions of the PFT equations (2-3) associated with the thermal equilibrium, therefore, they provide the exact relation for the chemical potential $\mu = const$ over all the domain of solution. Within the three-phase interface, this relation holds as well, while the concentration and phase distributions are clearly not uniform. Thus, it is natural to assume that the changes observed in the phase diagram obtained by solving PFT equations are due to the account of interfacial energy.

As is seen from the results obtained, the standard PFT model formulated above reproduces the general properties of phenomena associated with the formation of DZ and CM near the eutectic point, in particular, in the systems with amorphous Si. As for the experiments with crystalline Si films, they may reveal more complicated properties of the DZ [3]. Apparently, to describe these latter more adequately, the detailed description of component structures occuring in DZ is needed, eventually, by using the multi-phase version of PFT [14, 15]. At the same time, the basis for the consideration is the assumption is that the relatively wide DZ is associated with the equilibrium state of complete decomposition with high interfacial solid-solid energy, and the crucial for the model parameter, the DZ width, still remains of the same order.

The results of simulations suggest that the three-phase states and the associated properties of eutectic phase diagram would be rather common for eutectic systems. Another phenomenon, which can be viewed as the manifestation of the formation of three-phase-states, is the micro-separation of phases observed within the liquid eutectic above the eutectic point (see, for instance, the monographs [1, 16] and references therein). The experiments clearly demonstrate the strong deviation from the regular liquid structure in liquid eutectic even on macroscopic scales, especially, after the long-term (during several hours) keeping in isothermal conditions.

It is worth mentioning that, at present, the commonly accepted point of view is that the equilibrium state in binary eutectic below the eutectic temperature is associated with complete decomposition into ideal crystal phases (A + B) with sharp coherent or semi-coherent interface. Theoretical estimates of surface energy for such interface yield typically the values of $\Sigma \approx 1\,J/m^2$, which is three orders less than the surface energy of diffuse interface needed to provide the DZ width observed in experiments. The above consideration as well as the above cited experiments [2, 3] suggest that in actual fact the lowest free energy and the thermal equilibrium in eutectic would be associated with diffuse interface with irregular small-grained or amorphous structure, whereas the sharp interface between ideal crystal phases may turn out to be just a metastable state. A simple argument in favor of such a possibility is the obvious growth of disorder and the associated entropy contribution to the free energy in this case.

Of course, the PFT model used in this work is rather simplified. At the same time, the simulations demonstrate the role of the interfacial energy as the possible mechanism for contact melting and the above mentioned associated phenomena. To get a deeper insight into these issues, further experiments and theoretical studies (especially those based on the microscopic treatments) aimed at these questions are needed.

4. **Conclusions.**

To conclude, computer simulations of contact melting and associated interfacial phenomena in binary eutectic systems were performed on the basis of the standard phase-field model with miscibility gap in solid state. It is shown that the model predicts the existence of equilibrium three-phase (solid-liquid-solid) states above the eutectic temperature describing the complete decomposition with a liquid interface, which alternatively, can be viewed as phase separation in liquid eutectic. The results of simulations demonstrate the role of the interfacial energy as the possible mechanism for the phenomena of contact melting and formation of diffusion zone observed in the experiments with binary metal-silicon systems.

**Acknowledgements**
The authors thank Prof. O. Grigoriev for valuable discussions. The support of EOARD, Project No 118003 (STCU Project P-510) is acknowledged.

**Figure Captions**

Fig. 1. Schematic sketch of the experimental phase diagram of Cu-Si in the vicinity of the eutectic point [7]. The eutectic concentration is 0.3.

Fig. 2. Initial concentration profiles (solid lines 1 and 2) and phase distribution (dashed stochastic line) used in PFT simulations.

Fig. 3. Final steady-state profiles for phase (dashed line) and composition (solid lines) obtained in PFT simulations for the temperatures $T_{D1} = 923°K \, (650°C)$ (line 1) and $T_2 = 473°K \, (200°C)$ (line 2) for the time point $t = t_{max}$ (4000 sec).

Fig. 4. Final steady-state profiles for phase (dashed line) and composition (solid line) obtained in PFT simulations for the temperature $T_{D2} = 1083°K \, (810°C)$ (point D2).

Fig. 5. Average free energy density of the system as a function of time for the temperature $T_{D2} = 1083°K \, (810°C)$ and eutectic concentration $c = 0.3$ for the steady states: solid A + solid B (complete decomposition; solid line); solid A + liquid + solid B (complete decomposition with a liquid interface; dashed line).

Fig. 6. Average free energy density of the system vs. temperature for different steady states for the eutectic concentration c=0.3. The area in grey points to the temperature interval, where the three-phase state has the lowest free energy.

Fig. 7. Average free energy density of the system vs. temperature for different steady states for the concentration c=0.4. The areas in light and dark grey point to the temperature intervals of thermodynamical stability of three-phase 'solid A + liquid + solid B' and two-phase 'liquid + solid B' states, respectively.

Fig. 8. Phase diagram near the eutectic point obtained from PFT simulations.

**Figures**

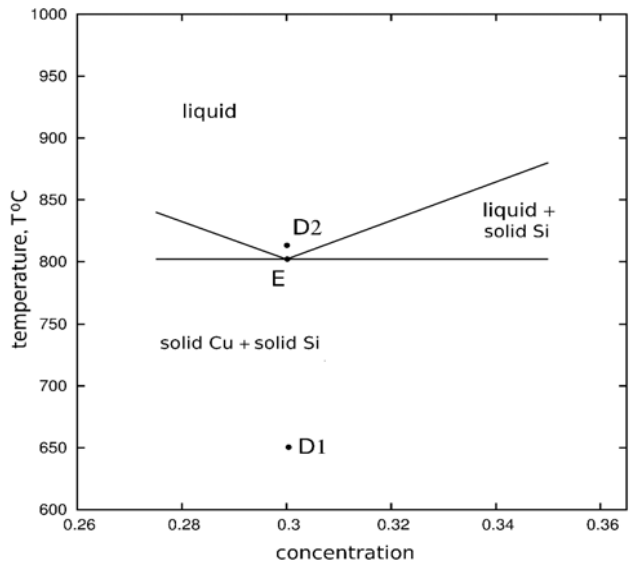

Fig. 1. Schematic sketch of the experimental phase diagram of Cu-Si in the vicinity of the eutectic point [7]. The eutectic concentration is 0.3.

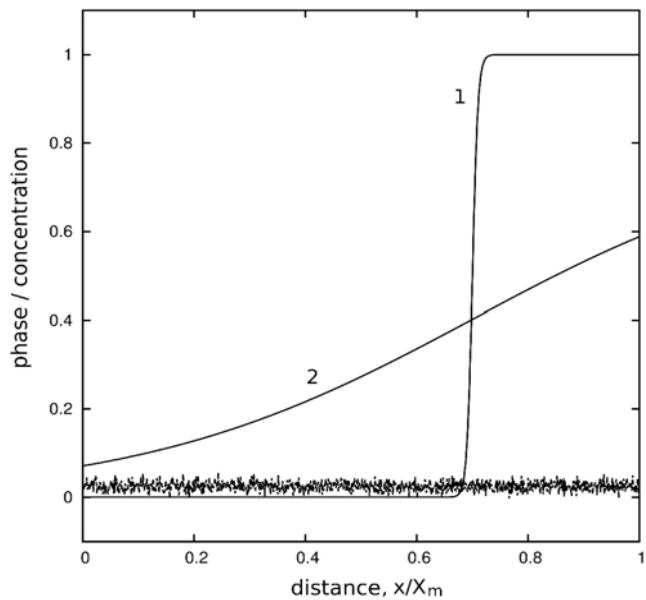

Fig. 2. Initial concentration profiles (solid lines 1 and 2) and phase distribution (dashed stochastic line) used in PFT simulations.

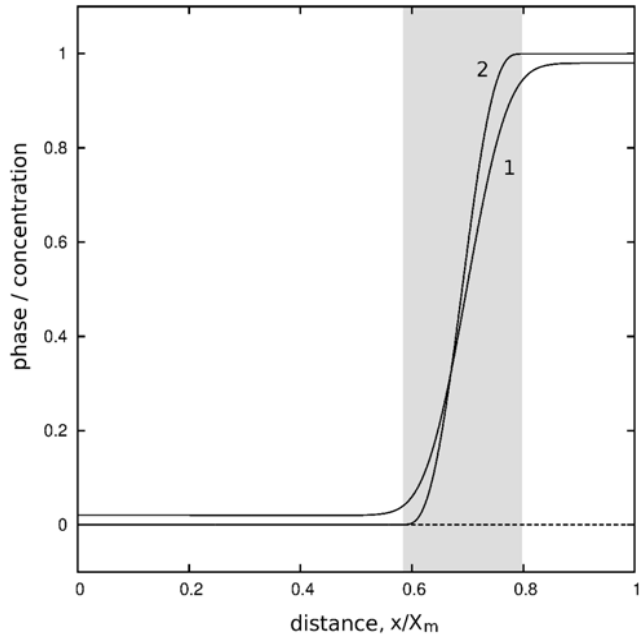

Fig. 3. Final steady-state profiles for phase (dashed line) and composition (solid lines) obtained in PFT simulations for the temperatures $T_{D1} = 923°K\ (650°C)$ (line 1) and $T_2 = 473°K\ (200°C)$ (line 2) for the time point $t = t_{max}$ (4000 sec).

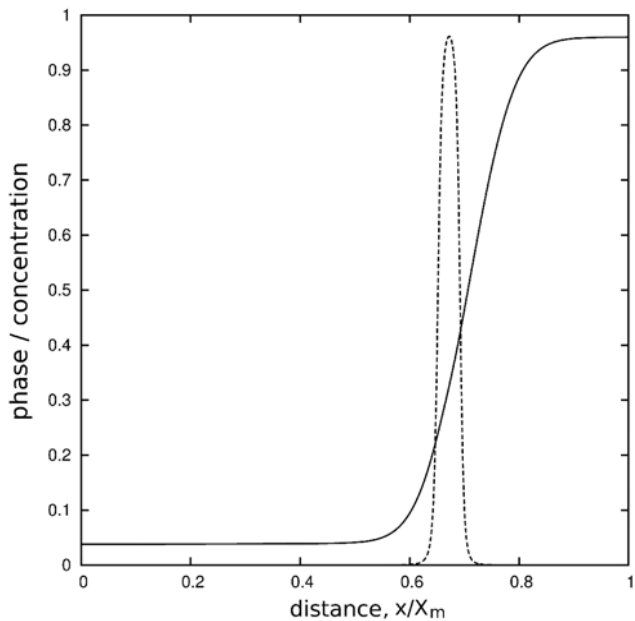

Fig. 4. Final steady-state profiles for phase (dashed line) and composition (solid line) obtained in PFT simulations for the temperature $T_{D2} = 1083°K\ (810°C)$ (point D2).

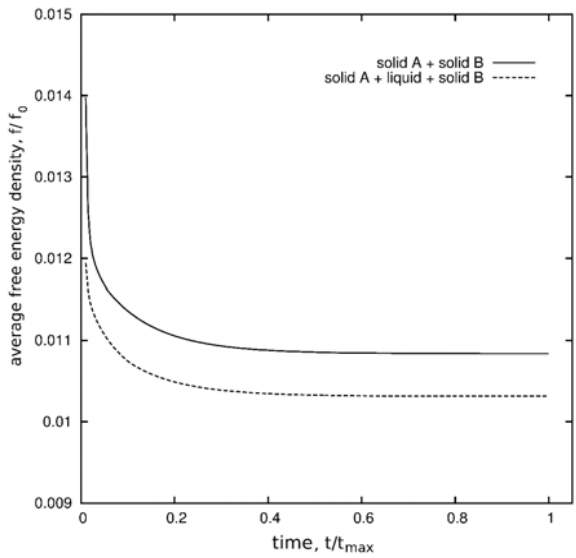

Fig. 5. Average free energy density of the system as a function of time for the temperature $T_{D2} = 1083°K\ (810°C)$ and eutectic concentration $c = 0.3$ for the steady states: solid A + solid B (complete decomposition; solid line); solid A + liquid + solid B (complete decomposition with a liquid interface; dashed line).

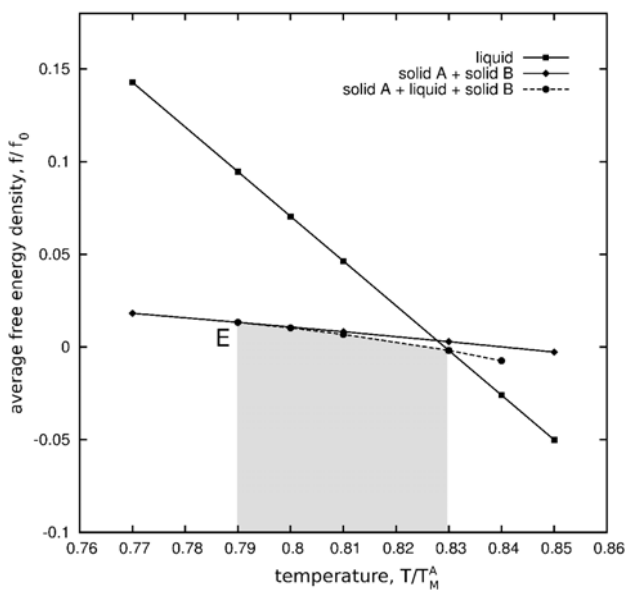

Fig. 6. Average free energy density of the system vs. temperature for different steady states for the eutectic concentration c=0.3. The area in grey points to the temperature interval, where the three-phase state has the lowest free energy.

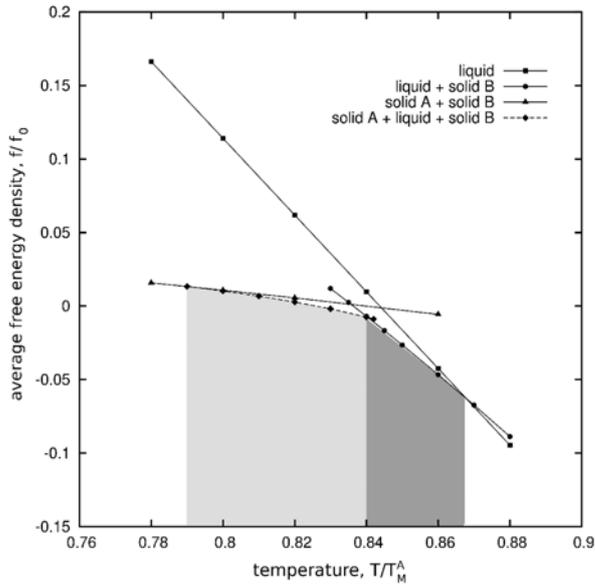

Fig. 7. Average free energy density of the system vs. temperature for different steady states for the concentration c=0.4. The areas in light and dark grey point to the temperature intervals of thermodynamical stability of three-phase 'solid A + liquid + solid B' and two-phase 'liquid + solid B' states, respectively.

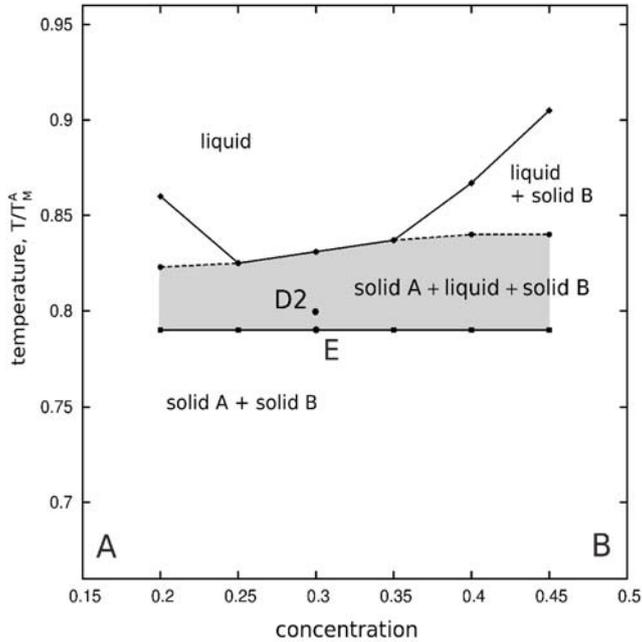

Fig. 8. Phase diagram near the eutectic point obtained from PFT simulations.